\documentclass[RNAAS]{aastex62}
\usepackage{graphicx}	
\usepackage{amsmath}	
\usepackage{amssymb}	
\usepackage{natbib}
\newcommand\xray{\hbox{X-ray}\space}
\newcommand\aox{$\alpha_{\rm ox}$\space}
\newcommand{\Lopt}{$L_{2500}$\space}
\newcommand{\heiiew}{\,\mbox{\ion{He}{2} EW}\space}
\begin{document}

\title{The $\alpha_{\rm ox}$--\ion{He}{2} EW connection in radio-loud quasars}

\correspondingauthor{John Timlin}
\affiliation{Department of Astronomy \& Astrophysics, 525 Davey Lab, The Pennsylvania State University, University Park, PA 16802, USA }
\email{jxt811@psu.edu}

\author{John Timlin}
\affiliation{Department of Astronomy \& Astrophysics, 525 Davey Lab, The Pennsylvania State University, University Park, PA 16802, USA }

\author{Shifu Zhu}
\affiliation{Department of Astronomy \& Astrophysics, 525 Davey Lab, The Pennsylvania State University, University Park, PA 16802, USA }

\author{W. N. Brandt}
\affiliation{Department of Astronomy \& Astrophysics, 525 Davey Lab, The Pennsylvania State University, University Park, PA 16802, USA }

\author{Ari Laor}
\affiliation{Physics Department, Technion, Haifa 32000, Israel }


\section{Abstract}

Radio-loud quasars (RLQs) are known to produce excess X-ray emission, compared to radio-quiet quasars (RQQs) of the same luminosity, commonly attributed to jet-related emission. Recently, we found that the \heiiew and \aox in RQQs are strongly correlated, which suggests that their extreme-ultraviolet (EUV) and \xray emission mechanisms are tightly related. Using 48 RLQs, we show that steep-spectrum radio quasars (SSRQs) and low radio-luminosity ($L_{\rm R}$) flat-spectrum radio quasars (FSRQs) follow the $\alpha_{\rm ox}$--\heiiew relation of RQQs. This suggests that the \xray and EUV emission mechanisms in these types of RLQs is the same as in RQQs, and is not jet related. High-$L_{\rm R}$ FSRQs show excess \xray emission given their \heiiew by a factor of $\approx$ 3.5, which suggests that only in this type of RLQ is the \xray production likely jet related. 

\section{The \ion{He}{2} and \aox properties of RLQs}

RLQs comprise $\approx$ 10--20\% of the total quasar population, are generally more \xray luminous than their RQQ counterparts, and typically exhibit a flatter \hbox{X-ray}-to-optical spectral slope ($\alpha_{\rm ox}$).\footnote{$\alpha_{\rm ox} \equiv \mathrm{log}(L_{\rm 2keV}/L_{\rm 2500})/\mathrm{log}(\nu_{\rm 2keV}/\nu_{\rm 2500})$} \citet{Zhu2020} (hereafter, Z20) studied the joint radio, optical, and \xray luminosities of a large, optically selected sample of typical RLQs and found evidence that their \xray emission largely originates in the corona, challenging the idea that a distinct jet component contributes substantially to the \xray emission of typical RLQs (e.g.,\ \citealt{Miller2011}). They suggested a small fraction ($<$10\%) of FSRQs (which have radio spectral slope $\alpha_r >-0.5$) may have a significant ($>$30\%) jet \xray contribution. In a follow-up investigation of the \xray spectral and variability properties of typical RLQs, they found that all but the most radio-luminous ($L_{\rm R} > 10^{34.3}\ \rm{erg\ s^{-1}\ Hz^{-1}}$; where $L_{\rm R}$ is the monochromatic 5~GHz luminosity) FSRQs had \xray emission as would be expected from the corona (Zhu et al. 2021, submitted; hereafter Z21). Here, we provide additional evidence supporting their result.

For this investigation, we follow Timlin et al. (2021, submitted; hereafter T21) who studied the $\alpha_{\rm ox}$--\heiiew relationship for RQQs. The \ion{He}{2} emission line is a weak, high-ionization line that provides a ``clean" measure of the number of EUV photons between 50--200 eV present (see Section~1 of T21). T21 found a tight correlation between \aox and \heiiew in RQQs, even after removing these parameters' \Lopt dependences, indicating that the EUV continuum is related to the \xray emission, and possibly originates in a ``warm Comptonization" coronal region. Given this tight correlation for RQQs, and that RLQ beamed jet emission is not expected to produce additional \ion{He}{2}, investigating the $\alpha_{\rm ox}$--\heiiew relation for RLQs might provide insight into the origin of RLQ \xray emission.

Our RLQ sample was constructed using RLQs from the three quasar samples outlined in Section~2 of T21 (\citealt{Just2007}; SDSS-RM, \citealt{Shen2019}; PG, \citealt{Schmidt1983}). These samples were selected due to their high-quality spectral coverage of \ion{He}{2}. In total, there are 13 RLQs from the T21 samples (1, 7, and 5 RLQs from \citealt{Just2007},  SDSS-RM, and PG, respectively). We also incorporated RLQs with $i<19$ and $1.6 \leq z \leq 3.5$ from Z20 into our RLQ sample, where the brightness cut ensured that we selected the highest-quality SDSS spectra, and the redshift range ensured that \ion{He}{2} is covered. After these cuts were imposed, 35 of the 729 quasars from Z20 were included in our sample. In total, our RLQ sample contains 48 RLQs,\footnote{see \url{https://doi.org/10.5281/zenodo.4688086}} of which there are 18 SSRQs ($\alpha_r \leq -0.5$), 21 low-$L_{\rm R}$ ($L_{\rm R} \leq 10^{34.3}\ \rm{erg\ s^{-1}\ Hz^{-1}}$) FSRQs, and 9 high-$L_{\rm R}$ FSRQs, where $\alpha_r$ and \heiiew were measured using the methods from Section~2.1.3 of Z20 and Section~3 of T21, respectively. The \aox values were either adopted from Z20 or computed following Section~2 of T21. 

\begin{figure}[h!]
\begin{center}
\includegraphics[scale=0.45,angle=0]{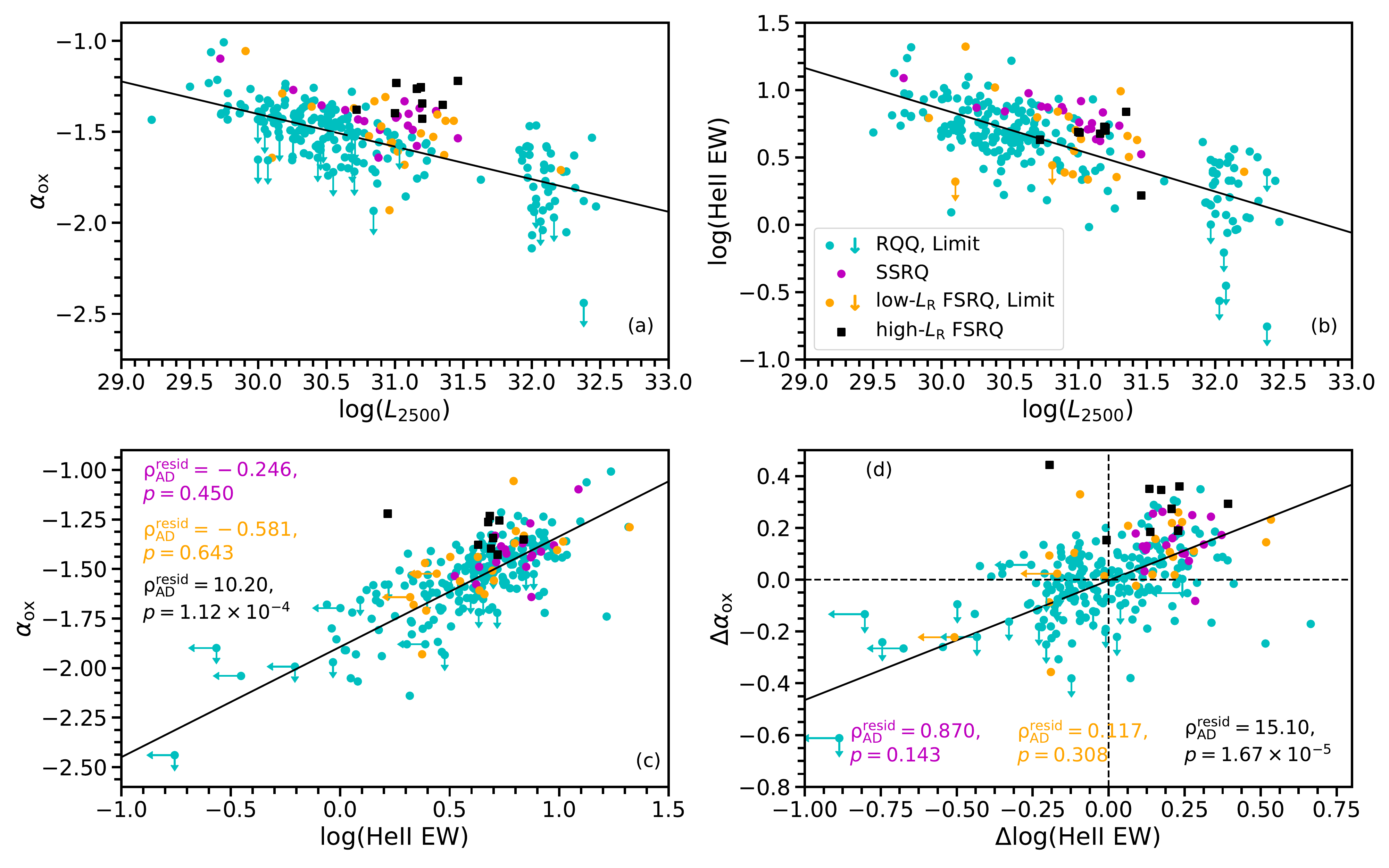}
\caption{Panel (a): $\alpha_{\rm ox}$ as a function of $L_{2500}$ for RQQs (light blue, from T21) and the SSRQs (purple points), low-$L_{\rm R}$ FSRQs (orange points), and high-$L_{\rm R}$ FSRQs (black squares). The solid black line shows the best-fit relationship in this parameter space for the RQQs (from T21). As expected, all three radio sub-samples generally have flatter \aox values than the RQQs at a given $L_{2500}$ indicating that the RLQs generally exhibit stronger X-ray emission than the RQQs.  Panel~(b): The \ion{He}{2} Baldwin effect for the RQQs and the RLQ sub-samples. As in the previous panel, all three RLQ sub-samples generally exhibit larger \heiiew than the RQQs for a given \Lopt value. Panel~(c): The relationship between \aox and \heiiew for both the RQQs and RLQ sub-samples. The \aox values of the SSRQs and the low-$L_{\rm R}$ FSRQs are well described by the best-fit relationship for the RQQs despite being both more \xray luminous and having a stronger \ion{He}{2} ionizing continuum, whereas the high-$L_{\rm R}$ FSRQs deviate significantly from this relationship. We quantify this using the distributions of the residuals about the best-fit line for the RQQs (black line). Anderson-Darling two-sample tests are performed between the RQQ residual distribution and the residual distributions of the three RLQ sub-samples (the results are listed in the Figure for each RLQ sub-sample by their respective color). The similarity between the residuals of the RQQs and the SSRQs and low-$L_{\rm R}$ FSRQs indicates that the \xray emission of these two RLQ populations largely originates in the corona; however, the RQQ and the high-$L_{\rm R}$ FSRQ residual distributions are not consistent, which suggests that a jet \xray component is likely present in these quasars (e.g.\ Z21). Panel~(d): Similar to panel (c), but we remove the \Lopt dependence of both \aox and \heiiew shown in panels (a) and (b). The AD tests of the residual distributions show a similar result as in panel (c). These panels together demonstrate that the \ion{He}{2} ionizing continuum is related to the coronal emission in both RLQs and RQQs even after removing the effects of $L_{2500}$, and that the \xray emission from typical RLQs largely originates in the corona. \label{fig:1}}
\end{center}
\end{figure}

Panels~(a) and (b) of Figure~\ref{fig:1} depict \aox and \heiiew as a function of \Lopt for the SSRQs, low-$L_{\rm R}$ FSRQs, and high-$L_{\rm R}$ FSRQs, respectively. The RQQs from T21 and the best-fit relation for the RQQs from T21 are also shown. As expected, the RLQs generally exhibit larger \aox values than the RQQs for a given $L_{2500}$. The RLQs also exhibit larger \heiiew in general than their RQQ counterparts indicating the presence of a stronger EUV continuum. Panel~(c) shows \aox as a function of \heiiew for the RLQs. Despite being more \xray luminous and having a stronger EUV continuum, the RLQs are generally consistent with the $\alpha_{\rm ox}$--\heiiew relation for RQQs. We use an Anderson-Darling two-sample test to determine if the distribution of the RQQ residuals (about the best-fit line) is consistent with those of the three RLQ sub-samples (see panel~(c)). Both the SSRQ and low-$L_{\rm R}$ FSRQ distributions are statistically consistent with the distribution of RQQs, while the high-$L_{\rm R}$ FSRQ sample is not. A similar conclusion is found after removing the \Lopt dependence of both \aox and \heiiew in panel~(d). These panels illustrate that the EUV continuum of SSRQs and low-$L_{\rm R}$ FSRQs strongly correlates with the \xray emission, and that the correlation is similar to the correlation for RQQs, consistent with the idea that both the \xray and EUV photons largely originate in a similar region as for RQQs. On the other hand, the likely beamed high-$L_{\rm R}$ FSRQs are not consistent with the RQQs in these parameter spaces, probably due to a significant jet \xray contribution as suggested in Z21. Finally, RLQs tend to have excess EUV emission compared to RQQs, at a given \Lopt (panel b), which may indicate that a more powerful accretion-disk corona is associated with jet formation.

\section{Supplementary Material}
\begin{figure}[h!]
\begin{center}
\includegraphics[scale=0.5,angle=0]{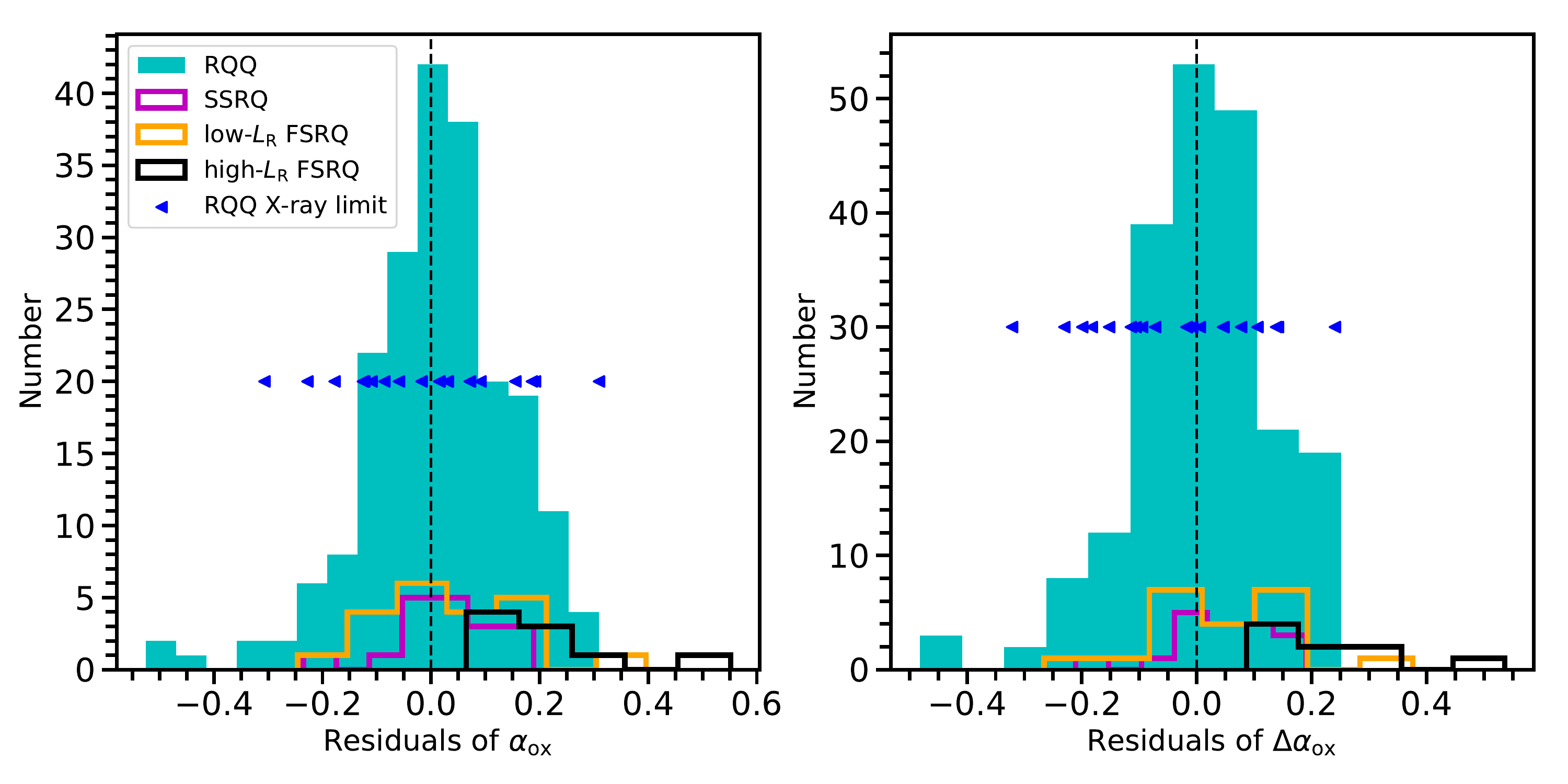}
\caption{The left-hand and right-hand panels depict the residual distributions from panels (c) and (d) in Figure~\ref{fig:1}, respectively. Clearly, in both panels, the high-$L_{\rm R}$ FSRQs are offset with respect to the other distributions. Note, all RLQs investigated are \xray detected. \label{fig:2}}
\end{center}
\end{figure}

In this Section, we provide some supplementary Figures and describe in more detail the columns in our data table. In Figure~\ref{fig:2}, we depict the residual distributions of the $\alpha_{\rm ox}$--\heiiew parameter space (left panel) and the $\Delta\alpha_{\rm ox}$--$\Delta$\heiiew parameter space (right panel). As presented above, the SSRQ and low-$L_{\rm R}$ FSRQ distributions are consistent with the RQQ distribution in these parameter spaces, and the high-$L_{\rm R}$ FSRQ distribution is clearly offset from zero residual in both cases. The mean of the high-$L_{\rm R}$ FSRQ residual distribution in the $\alpha_{\rm ox}$--\heiiew parameter space is 0.21, which indicates that the mean \xray luminosity of these quasars is $\approx 3.5$ times larger than expected from the \heiiew of the RQQs. The mean residual \aox values of the SSRQ and low-$L_{\rm R}$ FSRQ distributions are 0.03 and 0.04, respectively. All three RLQ sub-samples have a similar deviation about their mean of $\approx 0.12$; thus, the high-$L_{\rm R}$ FSRQ distribution is the only one not consistent with zero residual. 

\begin{figure}[h!]
\begin{center}
\includegraphics[scale=0.5,angle=0]{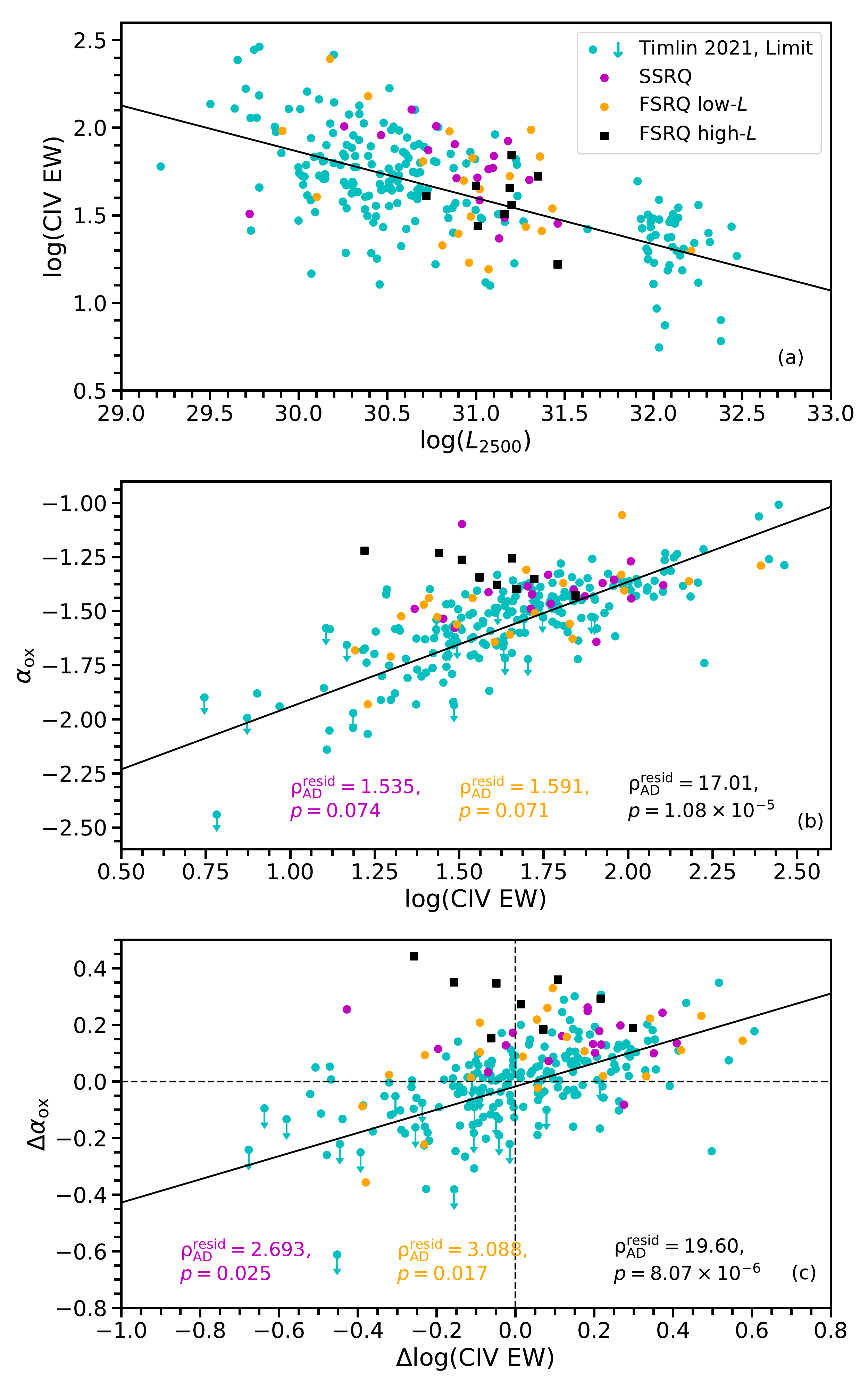}
\caption{Similar to panels (b)--(d) in Figure~\ref{fig:1}, however we now depict \ion{C}{4} EW instead of \ion{He}{2} EW. Panel~(b) demonstrates that the RLQs generally have a stronger \ion{C}{4} EW than the RQQs, however the offset is not as clear as for \ion{He}{2} EW. Panels~(c) and (d) also show that the SSRQs and the low-$L_{\rm R}$ FSRQs are slightly more offset from the RQQs in these parameter spaces than their \heiiew counterparts in Figure~\ref{fig:1}. As before, we quantify the consistency of the residual distributions of the RQQs with the SSRQs, low-$L_{\rm R}$ FSRQs, and high-$L_{\rm R}$ FSRQs using the Anderson-Darling two-sample test. We find that the residual distributions of the SSRQs and low-$L_{\rm R}$ FSRQs have a slightly lower $p$-value in this \ion{C}{4} EW parameter space than in the \heiiew parameter space, which may be due to the effects of the quasar environment on the \ion{C}{4} EW that do not materially affect the \ion{He}{2} EW. The high-$L_{\rm R}$ FSRQs remain strikingly inconsistent from the RQQs, indicating that their \xray emission likely has a jet-linked contribution.  \label{fig:3}}
\end{center}
\end{figure}

In Figure~\ref{fig:3} we reproduce panels (b)--(d) from Figure~\ref{fig:1}, however we now investigate the relationships with respect to \ion{C}{4} EW instead of \ion{He}{2} EW. In panel~(a) of Figure~\ref{fig:3} we find that the \ion{C}{4} EW for RLQs is generally larger than the expected value for RQQs at a given \Lopt as is the case with \ion{He}{2} EW, again indicating that the EUV emission in RLQs tends to be stronger than in RQQs. We depict the $\alpha_{\rm ox}$--\ion{C}{4} EW relation and the  $\Delta\alpha_{\rm ox}$--$\Delta$\ion{C}{4} EW relation in panels (b) and (c), respectively. In both panels~(b) and (c), the $\alpha_{\rm ox}$/$\Delta\alpha_{\rm ox}$ values of the SSRQ and low-$L_{\rm R}$ FSRQ populations appear to be slightly more offset at a given \ion{C}{4} EW than was found using \ion{He}{2} EW. As before, we use the Anderson-Darling two-sample test to compare the residual distributions of the RLQ populations with that of the RQQs. In both panels, the residual distributions of the high-$L_{\rm R}$ FSRQs still exhibit a more striking statistical difference from the distribution of the RQQs than the SSRQs and the low-$L_{\rm R}$ FSRQs, though the $p$-values for the latter two populations are smaller than what was found using \ion{He}{2}. These decreases may be due to the environmental effects of the quasar on \ion{C}{4} EW that do not significantly affect the \ion{He}{2} EW (see Section~1 of T21 for further discussion), though more data are required to perform a more detailed investigation. In both panels~(b) and (c), the spread of the SSRQs and low-$L_{\rm R}$ FSRQs are consistent with zero residual whereas the high-$L_{\rm R}$ FSRQs are still significantly offset.

Finally, we report the relevant multi-wavelength information for each of the 48 RLQs used in this work (see the associated Table). The first four columns of this table report the name of the sample from which the RLQ was drawn, the ID given to the quasar in that sample, and the RA and DEC (in decimal degrees) of each object. The redshift and the SDSS $i$-band magnitude (where applicable) are provided in columns~(5) and (6). Columns~(7)--(9) report the logarithm of the 2500~\AA , 5 GHz, and 2 keV luminosities (all in units of $\rm erg\ s^{-1}\ Hz^{-1}$). The logarithm of the radio-loudness parameter is provided in column (10) and the radio spectral slope in column (11).  Column~(12) reports the \aox value from either Z20, or computed following Section~2 of T21. Finally, in Columns~(13)--(15) we report whether the \ion{He}{2} emission line was detected in the spectrum, the logarithm of the rest-frame \heiiew (in \AA ), and the logarithm of the rest-frame \ion{C}{4} EW (in \AA ). 


\begin{thebibliography}{}

\bibitem[Schmidt \& Green(1983)]{Schmidt1983} Schmidt, M. \& Green, R.~F.\ 1983, \apj, 269, 352. doi:10.1086/161048

\bibitem[Kellermann et al.(1989)]{Kellerman1989} Kellermann, K.~I., Sramek, R., Schmidt, M., et al.\ 1989, \aj, 98, 1195. doi:10.1086/115207

\bibitem[Just et al.(2007)]{Just2007} Just, D.~W., Brandt, W.~N., Shemmer, O., et al.\ 2007, \apj, 665, 1004. doi:10.1086/519990

\bibitem[Miller et al.(2011)]{Miller2011} Miller, B.~P., Brandt, W.~N., Schneider, D.~P., et al.\ 2011, \apj, 726, 20. doi:10.1088/0004-637X/726/1/20


\bibitem[Shen et al.(2019)]{Shen2019} Shen, Y., Hall, P.~B., Horne, K., et al.\ 2019, \apjs, 241, 34. doi:10.3847/1538-4365/ab074f

\bibitem[Timlin et al. (2021)]{Timlin2021} Timlin, J.~D., Brandt, W.~N., Laor, A.\ 2021, \mnras, Manuscript submitted for publication. (T21)

\bibitem[Zhu et al.(2020)]{Zhu2020} Zhu, S.~F., Brandt, W.~N., Luo, B., et al.\ 2020, \mnras, 496, 245. doi:10.1093/mnras/staa1411 (Z20)

\bibitem[Zhu et al.(2021)]{Zhu2021} Zhu, S.~F.,  Timlin, J.~D., Brandt, W.~N.\ 2021, \mnras, Manuscript submitted for publication. (Z21)

\end{thebibliography}
\end{document}